\documentclass{epl}

\title{Two-dimensional finite-difference lattice Boltzmann method for the complete
Navier-Stokes equations of binary fluids}
\shorttitle{2-d FDLBM for Navier-Stokes equations}
\author{Aiguo Xu}
\institute{Department of Physics, Yoshida-South Campus, Kyoto University, \\
Sakyo-ku, Kyoto, 606-8501, Japan }

\begin{document}
\maketitle

\begin{abstract}
Based on Sirovich's two-fluid kinetic theory and a dodecagonal discrete
velocity model, a two-dimensional 61-velocity finite-difference lattice
Boltzmann method for the complete Navier-Stokes equations of binary fluids
is formulated. Previous constraints, in most existing lattice Boltzmann
methods, on the studied systems, like isothermal and nearly incompressible,
are released within the present method. This method is
designed to simulate compressible and thermal binary fluid mixtures. The
validity of the proposed method is verified by investigating (i) the Couette
flow and (ii) the uniform relaxation process of the two components.
\end{abstract}

\section{Introduction}

Lattice Boltzmann Method (LBM) has become a viable and promising numerical
scheme for simulating fluid flows. There are several options to discretize the
Boltzmann equation: (i) Standard LBM (SLBM)\cite{Succi};
(ii) Finite-Difference LBM (FDLBM)\cite{Succi,FDLBM,SofoneaPRE};
(iii) Finite-Volume LBM\cite{Succi,FVLBM};
(iv) Finite-Element LBM\cite{Succi,FELBM}; etc. 
These kinds of schemes are expected to be complementary in the
LBM studies.

Even though various LBMs for multicomponent fluids\cite
{PRE6635301,PRE6835302,PRE6756105,PRA434320,PRE474247,PFA52557,JSP81379,EPL32463,PRL83576,IJCES373,ICCS2003,CEJP2382,ShanChen,Coveney}
have been proposed and developed , (i) most existing methods belong to the
SLBM\cite
{PRE6635301,PRE6835302,PRE6756105,PRA434320,PRE474247,PFA52557,JSP81379,EPL32463,PRL83576,IJCES373,ICCS2003}%
, and/or based on the single-fluid theory\cite
{PRE6835302,PRE6756105,PRA434320,PRE474247,PFA52557,JSP81379,EPL32463,PRL83576,ICCS2003,CEJP2382,PhysA299494}%
; (ii) in Ref. \cite{PRE6635301} a SLBM based on Sirovich's two-fluid
kinetic theory\cite{PF5908} is proposed; (iii) nearly all the studies are focused on isothermal and nearly
incompressible systems. In a recent study\cite{XuFDLBM1}, Sirovich's kinetic
theory is clarified and corresponding two-fluid FDLBMs for Euler equations
and isothermal Navier-Stokes equations are presented. In this letter we
propose a two-fluid FDLBM for the complete Navier-Stokes equations,
including the energy equation.

\section{Formulation and verification of the FDLBM}

The formulation of a FDLBM consists of three steps: (i) select or design an
appropriate discrete velocity model (DVM), (ii) formulate the discrete
local equilibrium distribution function, (iii) choose a
finite-difference scheme.
The continuous Boltzmann equation has infinite velocities, so the rotational
invariance is automatically satisfied. Recovering rotational invariant
macroscopic equations from a discrete-finite-velocity microscopic dynamics
imposes constraints on the isotropy of DVM and the finite-difference
scheme used. In this Letter, the
proposed FDLBM is based on the following DVM, 
\begin{equation}
v_0=0\mathtt{,}\,\,\mathbf{v}_{ki}=v_k\left[ \cos \left( \frac{i\pi }%
6\right) \mathtt{,}\sin \left( \frac{i\pi }6\right) \right] \mathtt{,}i=1%
\mathtt{,}2\mathtt{,}\cdots \mathtt{,}12\mathtt{,}  \label{DVM1}
\end{equation}
where $k$ indicates the $k$-th group of particle velocities and $i$
indicates the direction of the particle speed. It is easy find that (i) its
odd rank tensors are zero, and (ii) its initial four even rank tensors
satisfy 
\begin{equation}
\begin{array}{l}
\sum_{i=1}^{12}v_{ki\alpha }v_{ki\beta }=6v_k^2\delta _{\alpha \beta }%
\mathtt{,}\,\,\sum_{i=1}^{12}v_{ki\alpha }v_{ki\beta }v_{ki\gamma
}v_{ki\delta }=\frac 32v_k^4\Delta _{\alpha \beta \gamma \delta }\mathtt{,}
\\ 
\sum_{i=1}^{12}v_{ki\alpha }v_{ki\beta }v_{ki\gamma }v_{ki\delta }v_{ki\mu
}v_{ki\nu }=\frac 14v_k^6\Delta _{\alpha \beta \gamma \delta \mu \nu }%
\mathtt{,} \\ 
\sum_{i=1}^{12}v_{ki\alpha }v_{ki\beta }v_{ki\gamma }v_{ki\delta }v_{ki\mu
}v_{ki\nu }v_{ki\lambda }v_{ki\pi }=\frac 1{32}v_k^8\Delta _{\alpha \beta
\gamma \delta \mu \nu \lambda \pi }\mathtt{,}
\end{array}
\label{DVM2}
\end{equation}
where $\alpha $, $\beta $, $\cdots $ indicate $x$ or $y$ component and 
\begin{equation}
\Delta _{\alpha \beta \gamma \delta }=\delta _{\alpha \beta }\delta _{\gamma
\delta }+\delta _{\alpha \gamma }\delta _{\beta \delta }+\delta _{\alpha
\delta }\delta _{\beta \gamma }\mathtt{,}  \label{DVM7}
\end{equation}
\begin{equation}
\Delta _{\alpha \beta \gamma \delta \mu \nu }=\delta _{\alpha \beta }\Delta
_{\gamma \delta \mu \nu }+\delta _{\alpha \gamma }\Delta _{\beta \delta \mu
\nu }+\delta _{\alpha \delta }\Delta _{\beta \gamma \mu \nu }+\delta
_{\alpha \mu }\Delta _{\beta \gamma \delta \nu }+\delta _{\alpha \nu }\Delta
_{\beta \gamma \delta \mu }\mathtt{,}  \label{DVM8}
\end{equation}
\begin{eqnarray}
\Delta _{\alpha \beta \gamma \delta \mu \nu \lambda \pi } &=&\delta _{\alpha
\beta }\Delta _{\gamma \delta \mu \nu \lambda \pi }+\delta _{\alpha \gamma
}\Delta _{\beta \delta \mu \nu \lambda \pi }+\delta _{\alpha \delta }\Delta
_{\beta \gamma \mu \nu \lambda \pi }+\delta _{\alpha \mu }\Delta _{\beta
\gamma \delta \nu \lambda \pi }  \nonumber \\
&&+\delta _{\alpha \nu }\Delta _{\beta \gamma \delta \mu \lambda \pi
}+\delta _{\alpha \lambda }\Delta _{\beta \gamma \delta \mu \nu \pi }+\delta
_{\alpha \pi }\Delta _{\beta \gamma \delta \mu \nu \lambda }\mathtt{.}
\label{DVM9}
\end{eqnarray}
It is clear that this DVM is isotropic up to, at least, its $9$th rank
tensor.

We consider a binary mixture with two components, $A$ and $B$, where the
masses and temperatures of the two components are not significantly
different. The interparticle collisions can be divided into two kinds:
collisions within the same species (self-collision) and collisions among
different species (cross-collision)\cite{PF5908}. Based on the DVM (\ref
{DVM1}), the $2$-dimensional BGK\cite{PR94511} kinetic equation for species $%
A$ reads\cite{XuFDLBM1}, 
\begin{equation}
\partial _tf_{ki}^A+\mathbf{v}_{ki}^A\cdot \frac \partial {\partial \mathbf{r%
}}f_{ki}^A-\mathbf{a}^A\cdot \frac{\left( \mathbf{v}_{ki}^A-\mathbf{u}%
^A\right) }{\Theta ^A}f_{ki}^{A\left( 0\right) }=J_{ki}^{AA}+J_{ki}^{AB}
\label{be1}
\end{equation}
where 
\begin{equation}
J_{ki}^{AA}=-\left[ f_{ki}^A-f_{ki}^{A(0)}\right] /\tau ^{AA}\mathtt{\ ,}%
\,\,J_{ki}^{AB}=-\left[ f_{ki}^A-f_{ki}^{AB(0)}\right] /\tau ^{AB}
\label{be2}
\end{equation}
\begin{equation}
f_{ki}^{A(0)}=\frac{n^A}{2\pi \Theta ^A}\exp \left[ -\frac{\left( \mathbf{v}%
_{ki}^A-\mathbf{u}^A\right) ^2}{2\Theta ^A}\right] \mathtt{,}%
\,\,f_{ki}^{AB(0)}=\frac{n^A}{2\pi \Theta ^{AB}}\exp \left[ -\frac{\left( 
\mathbf{v}_{ki}^A-\mathbf{u}^{AB}\right) ^2}{2\Theta ^{AB}}\right] 
\label{be3}
\end{equation}
\begin{equation}
\Theta ^A=k_BT^A/m^A\mathtt{,}\,\,\Theta ^{AB}=k_BT^{AB}/m^A  \label{be4}
\end{equation}
$f^{A(0)}$ and $f^{AB(0)}$ are the corresponding Maxwellian distribution
functions. $n^A$, $\mathbf{u}^A$, $T^A$ are the local density, hydrodynamic
velocity and temperature of species $A$. $\mathbf{u}^{AB}$, $T^{AB}$ are the
hydrodynamic velocity and temperature of the mixture after equilibration
process. $\mathbf{a}^A$ is the acceleration of species $A$ due to the
effective external field.

For species $A$, we have 
\begin{equation}
n^A=\sum_{ki}f_{ki}^A\mathtt{\ ,}\,\,n^A\mathbf{u}^A=\sum_{ki}\mathbf{v}%
_{ki}^Af_{ki}^A\mathtt{,}\,\,P^A\mathtt{(}e_{\mathtt{int}}^A=n^Ak_BT^A%
\mathtt{)}=\sum_{ki}\frac 12m^A\mathbf{(v}_{ki}^A-\mathbf{u}^A\mathbf{)}%
^2f_{ki}^A  \label{be5}
\end{equation}
where $P^A$ ($e_{\mathtt{int}}^A$) is the local pressure (internal mean
kinetic energy).
For species $B$, we have similar relations. For the mixture, we have 
\begin{equation}
\mathbf{u}^{AB}=\left( \rho ^A\mathbf{u}^A+\rho ^B\mathbf{u}^B\right) /\rho 
\mathtt{,}\,\,nk_BT^{AB}=\sum_{ki}\frac 12\left[ \mathbf{(v}_{ki}^A-\mathbf{u%
}^{AB}\mathbf{)}^2m^Af_{ki}^A+\mathbf{(v}_{ki}^B-\mathbf{u}^{BA}\mathbf{)}%
^2m^Bf_{ki}^B\right]  \label{be6}
\end{equation}
where $\rho ^A=n^Am^A$, $n=n^A+n^B$ and $\rho =\rho ^A+\rho ^B$. Three sets
of hydrodynamic quantities (for the two components $A$, $B$ and for the
mixture) are involved, but only two sets of them are independent. So this is
a two-fluid model. Without lossing generality, we focus on hydrodynamics of
the two individual species. By expanding the local equilibrium distribution
function $f^{AB(0)}$ around $f^{A(0)}$ to the first order in flow velocity
and temperature, the BGK model (\ref{be1}-\ref{be4}) becomes 
\begin{equation}
\partial _tf_{ki}^A+\mathbf{v}_{ki}^A\cdot \frac \partial {\partial \mathbf{r%
}}f_{ki}^A-\mathbf{a}^A\cdot \frac{\left( \mathbf{v}_{ki}^A-\mathbf{u}%
^A\right) }{\Theta ^A}f_{ki}^{A\left( 0\right) }=Q_{ki}^{AA}+Q_{ki}^{AB}
\label{be8}
\end{equation}
\begin{equation}
Q_{ki}^{AA}=-\left( \frac 1{\tau ^{AA}}+\frac 1{\tau ^{AB}}\right) \left[
f_{ki}^A-f_{ki}^{A(0)}\right]  \label{be9}
\end{equation}
\begin{eqnarray}
Q_{ki}^{AB} &=&-\frac{f_{ki}^{A(0)}}{\rho ^A\Theta ^A}\{\mu _D^A\left( 
\mathbf{v}_{ki}^A-\mathbf{u}^A\right) \cdot (\mathbf{u}^A-\mathbf{u}^B) 
\nonumber \\
&&+\mu _T^A\left[ \frac{\left( \mathbf{v}_{ki}^A-\mathbf{u}^A\right) ^2}{%
2\Theta ^A}-1\right] (T^A-T^B)-M^A\left[ \frac{\left( \mathbf{v}_{ki}^A-%
\mathbf{u}^A\right) ^2}{2\Theta ^A}-1\right] (\mathbf{u}^A-\mathbf{u}^B)^2\}
\label{be11}
\end{eqnarray}
where $\mu _D^A=\rho ^A\rho ^B/(\tau ^{AB}\rho )$, $\mu _T^A=k_Bn^An^B/(\tau
^{AB}n)$, $M^A=n^A\rho ^A\rho ^B/(2\tau ^{AB}n\rho )$.

Now, we go to the second step: formulate $f_{ki}^{A\left( 0\right) }$. The
 continuous Maxwellian $f^{A\left( 0\right) }$ possesses an infinite sequence
of moment properties. The Chapman-Enskog analysis\cite{Chap} shows that, requiring
the discrete $f_{ki}^{A\left( 0\right) }$ to follow the initial eight ones
is sufficient to describe the same Navier-Stokes equations, 
\begin{equation}
\frac{\partial \rho ^A}{\partial t}+\frac \partial {\partial r_\alpha
}\left( \rho ^Au_\alpha ^A\right) =0,  \label{mass}
\end{equation}
\begin{eqnarray}
&\frac \partial {\partial t}\left( \rho ^A u_\alpha ^A\right) + \frac
\partial {\partial r_\beta }\left( \rho ^A u_\alpha ^Au_\beta ^A\right) + 
\frac{\partial P^A}{\partial r_\alpha }-\rho ^Aa_\alpha ^A-\frac \partial
{\partial r_\beta }\left[ \eta ^A\left( \frac{\partial u_\alpha ^A}{\partial
r_\beta }+\frac{\partial u_\beta ^A}{\partial r_\alpha }-\frac{\partial
u_\gamma ^A}{\partial r_\gamma }\delta _{\alpha \beta }\right) \right] 
\nonumber \\
&+\frac{\rho ^A\rho ^B}{\tau ^{AB}\rho }\left( u_\alpha ^A-u_\alpha
^B\right) =0 \mathtt{,} \label{momentum}
\end{eqnarray}
\begin{eqnarray}
&\frac{\partial e^A}{\partial t}\mathbf{+}\frac \partial {\partial r_\alpha
}\left[ \left( e^A+P^A\right) u_\alpha ^A\right] -\rho ^A\mathbf{a}^A\cdot 
\mathbf{u}^A-\frac \partial {\partial r_\alpha }\left[ k^A\frac{\partial
\left( k_BT^A\right) }{\partial r_\alpha }+\eta ^Au_\beta ^A\left( \frac{%
\partial u_\alpha ^A}{\partial r_\beta }+\frac{\partial u_\beta ^A}{\partial
r_\alpha }-\frac{\partial u_\gamma ^A}{\partial r_\gamma }\delta _{\alpha
\beta }\right) \right]  \nonumber \\
&+\frac{\rho ^A\rho ^B}{\tau ^{AB}\rho }\left[ \left( u^A\right) ^2-\mathbf{u%
}^A\cdot \mathbf{u}^B\right] +\frac{n^An^B}{\tau ^{AB}n}k_B\left(
T^A-T^B\right) -n^A\frac{\rho ^A\rho ^B}{2\tau ^{AB}n\rho }\left( \mathbf{u}%
^A-\mathbf{u}^B\right) ^2=0 \mathtt{,} \label{energy}
\end{eqnarray}
where 
\begin{equation}
e^A=e_{\mathtt{int}}^A+\frac 12\rho ^A\left( u^A\right) ^2\mathtt{,}\,\,\eta
^A=P^A\tau ^{AA}\tau ^{AB}/\left( \tau ^{AA}+\tau ^{AB}\right) \mathtt{,}%
\,\,k^A=2n^A\Theta ^A\tau ^{AA}\tau ^{AB}/\left( \tau ^{AA}+\tau
^{AB}\right) \mathtt{.}  \label{coefficient}
\end{equation}
Recall that $\mathbf{u}^A$ ($\mathbf{u}^B$) is a small quantity. By
using Eq. (\ref{mass}),  $P^A = n^A k_B T^A$, and  neglecting the second 
and higher order terms in $\mathbf{u}^A$,
Eq. (\ref{momentum}) shows that the diffusion velocity, 
$u^B_\alpha -u^A_\alpha$, is related to the gradients of $n^A$ and $T^A$.

The first three requirements on $f_{ki}^{A(0)}$ are referred to Eq. (\ref{be5}) with $f_{ki}^A$
replaced by $f_{ki}^{A(0)}$, and the remaining five are 
\begin{equation}
\sum_{ki}m^Av_{ki\alpha }^Av_{ki\beta }^Af_{ki}^{A(0)}=P^A\delta _{\alpha
\beta }+\rho ^Au_\alpha ^Au_\beta ^A  \label{rt4}
\end{equation}
\begin{equation}
\sum_{ki}m^Av_{ki\alpha }^Av_{ki\beta }^Av_{ki\gamma
}^Af_{ki}^{A(0)}=P^A\left( u_\gamma ^A\delta _{\alpha \beta }+u_\alpha
^A\delta _{\beta \gamma }+u_\beta ^A\delta _{\gamma \alpha }\right) +\rho
^Au_\alpha ^Au_\beta ^Au_\gamma ^A  \label{rt5}
\end{equation}
\begin{equation}
\sum_{ki}\frac 12m^A\left( v_{ki}^A\right) ^2v_{ki\alpha
}^Af_{ki}^{A(0)}=2n^Ak_BT^Au_\alpha ^A+\frac 12\rho ^A\left( u^A\right)
^2u_\alpha ^A  \label{rt6}
\end{equation}
\begin{eqnarray}
&&\sum_{ki}\frac 12m^A\left( v_{ki}^A\right) ^2v_{ki\alpha }^Av_{ki\beta
}^Af_{ki}^{A(0)}=2P^A\Theta ^A\delta _{\alpha \beta }+\frac 12P^A\left(
u^A\right) ^2\delta _{\alpha \beta }  \nonumber \\
&&+3P^Au_\alpha ^Au_\beta ^A+\frac 12\rho ^A\left( u^A\right) ^2u_\alpha
^Au_\beta ^A  \label{rt7}
\end{eqnarray}
\begin{equation}
\sum_{ki}\frac 12m^A\left( v_{ki}^A\right) ^4v_{ki\alpha
}^Af_{ki}^{A(0)}=\left[ 12P^A\Theta ^A+6P^A\left( u^A\right) ^2+\frac
12\rho ^A\left( u^A\right) ^4\right] u_\alpha ^A  \label{r8}
\end{equation}

The requirement equation (\ref{r8}) contains the fifth order of the flow
velocity $\mathbf{u}^A$. So it is sufficient to expand $f_{ki}^{A(0)}$ in
polynomial up to the fifth order of $\mathbf{u}^A$: 
\begin{eqnarray}
f_{ki}^{A(0)} &=&n^AF_k^A\left\{ \left[ 1-\frac{(u^A)^2}{2\Theta ^A}+\frac{%
(u^A)^4}{8\left( \Theta ^A\right) ^2}\right] +\frac{v_{ki\xi }^Au_\xi ^A}{%
\Theta ^A}\left[ 1-\frac{(u^A)^2}{2\Theta ^A}+\frac{(u^A)^4}{8\left( \Theta
^A\right) ^2}\right] \right.  \nonumber \\
&&+\frac{v_{ki\xi }^Av_{ki\pi }^Au_\xi ^Au_\pi ^A}{2\left( \Theta ^A\right)
^2}\left[ 1-\frac{(u^A)^2}{2\Theta ^A}\right] +\frac{v_{ki\xi }^Av_{ki\pi
}^Av_{ki\eta }^Au_\xi ^Au_\pi ^Au_\eta ^A}{6\left( \Theta ^A\right) ^3}%
\left[ 1-\frac{(u^A)^2}{2\Theta ^A}\right]  \nonumber \\
&&\left. +\frac{v_{ki\xi }^Av_{ki\pi }^Av_{ki\eta }^Av_{ki\lambda }^Au_\xi
^Au_\pi ^Au_\eta ^Au_\lambda ^A}{24\left( \Theta ^A\right) ^4}+\frac{%
v_{ki\xi }^Av_{ki\pi }^Av_{ki\eta }^Av_{ki\lambda }^Av_{ki\delta }^Au_\xi
^Au_\pi ^Au_\eta ^Au_\lambda ^Au_\delta ^A}{120\left( \Theta ^A\right) ^5}%
\right\}  \nonumber \\
&&+\cdots  \label{d1}
\end{eqnarray}
where 
\begin{equation}
F_k^A=\frac 1{2\pi \Theta ^A}\exp \left[ -\frac{(v_k^A)^2}{2\Theta ^A}%
\right] \mathtt{.}  \label{d2}
\end{equation}
The truncated equilibrium distribution function $f_{ki}^{A(0)}$ (\ref{d1})
contains the fifth rank tensor of the particle velocity $\mathbf{v}^A$ and
the requirement (\ref{rt5}) contains its third rank tensor. Thus, a DVM
being isotropic up to its $8$th rank tensors is enough to recover the
physical isotropy of the continuous Boltzmann equations to the Navier-Stokes
level. So DVM (\ref{DVM1}) is an appropriate choice. To calculate the
discrete $f_{ki}^{A(0)}$, one first needs calculate the factor $F_k^A$. $%
F_k^A$ is determined by the eight requirements on $f_{ki}^{A(0)}$ and the
isotropic properties of the DVM (\ref{DVM1}). Following the same procedure
as described in \cite{XuFDLBM1}, we obtain 
\begin{eqnarray}
\sum_{ki}F_k^A &=&1\mathtt{,}\,\,\sum_kF_k^A\left( v_k^A\right) ^2=\frac{%
\Theta ^A}6\mathtt{,}\,\,\sum_kF_k^A\left( v_k^A\right) ^4=\frac 23\left(
\Theta ^A\right) ^2\mathtt{,}  \nonumber \\
\sum_kF_k^A\left( v_k^A\right) ^6 &=&4\left( \Theta ^A\right) ^3\mathtt{,}%
\,\,\sum_kF_k^A\left( v_k^A\right) ^8=32\left( \Theta ^A\right) ^4\mathtt{,}%
\,\,\sum_kF_k^A\left( v_k^A\right) ^{10}=320\left( \Theta ^A\right) ^5
\label{re}
\end{eqnarray}
Once a zero speed, $v_0^A=0$, and other five nonzero ones, $v_k^A$ ($k=1$, $%
2 $, $3$, $4$, $5$) are chosen, $F_k^A$ ($k=0$, $1$, $2$, $3$, $4$, $5$)
will be fixed.

We come to the third step: finite-difference implementation of the discrete
kinetic method. There are several choices\cite{CEJP2382} available.
One possibility is shown below, 
\begin{equation}
f_{ki}^{A,(n+1)}=f_{ki}^{A,(n)}+\left[ \mathbf{a}^A\cdot \frac{\left( 
\mathbf{v}_{ki}^A-\mathbf{u}^A\right) }{\Theta ^A}f_{ki}^{A\left( 0\right)
}+Q_{ki}^{AA,(n)}+Q_{ki}^{AB,(n)}-\mathbf{v}_{ki}^A\cdot \frac{\partial
f_{ki}^{A,(n)}}{\partial \mathbf{r}}\right] \Delta t\mathtt{,}  \label{fd1}
\end{equation}
where the second superscripts $n$, $n+1$ indicate the consecutive two
iteration steps, $\Delta t$ the time step; the spatial derivatives are
calculated as 
\begin{equation}
\frac{\partial f_{ki}^{A,(n)}}{\partial \alpha }=\left\{ 
\begin{array}{ll}
(3f_{ki,I}^{A,(n)}-4f_{ki,I-1}^{A,(n)}+f_{ki,I-2}^{A,(n)})/(2\Delta \alpha )
& \mathtt{if}\,v_{ki\alpha }^A\geq 0 \\ 
(3f_{ki,I}^{A,(n)}-4f_{ki,I+1}^{A,(n)}+f_{ki,I+2}^{A,(n)})/(-2\Delta \alpha )
& \mathtt{if}\,v_{ki\alpha }^A<0
\end{array}
\right. \mathtt{,}  \label{fd2}
\end{equation}
where $\alpha =x,y$, the third subscripts $I-2$, $I-1$, $I$, $I+1$, $I+2$
indicate consecutive mesh nodes in the $\alpha $ direction.

The validity of the formulated FDLBM is verified through two test
examples. (The Boltzmann constant $k_B=1$.) The first one is the isothermal and incompressible Couette flow
with a single component. In this case, $A=B$. The initial state of the fluid
is static. The distance between the two walls is $D$. At time $t=0$ they
start to move at velocities $U$, $-U$, respectively. The horizontal velocity
profiles of species $A$ or $B$ along a vertical line agree with the
following analytical solution, 
\begin{equation}
u=\gamma y-\sum_j(-1)^{j+1}\frac{\gamma D}{j\pi }\exp (-\frac{4j^2\pi ^2\eta 
}{\rho D^2}t)\sin (\frac{2j\pi }Dy)\mathtt{,}  \label{uy}
\end{equation}
where $\gamma =2U/D$ is the imposed shear rate, $j$ is an integer, the
two walls locate at $y=\pm D/2$. (For example, see Fig. 1.)

\begin{figure}[tbp]
\twofigures[height=3.7 cm]{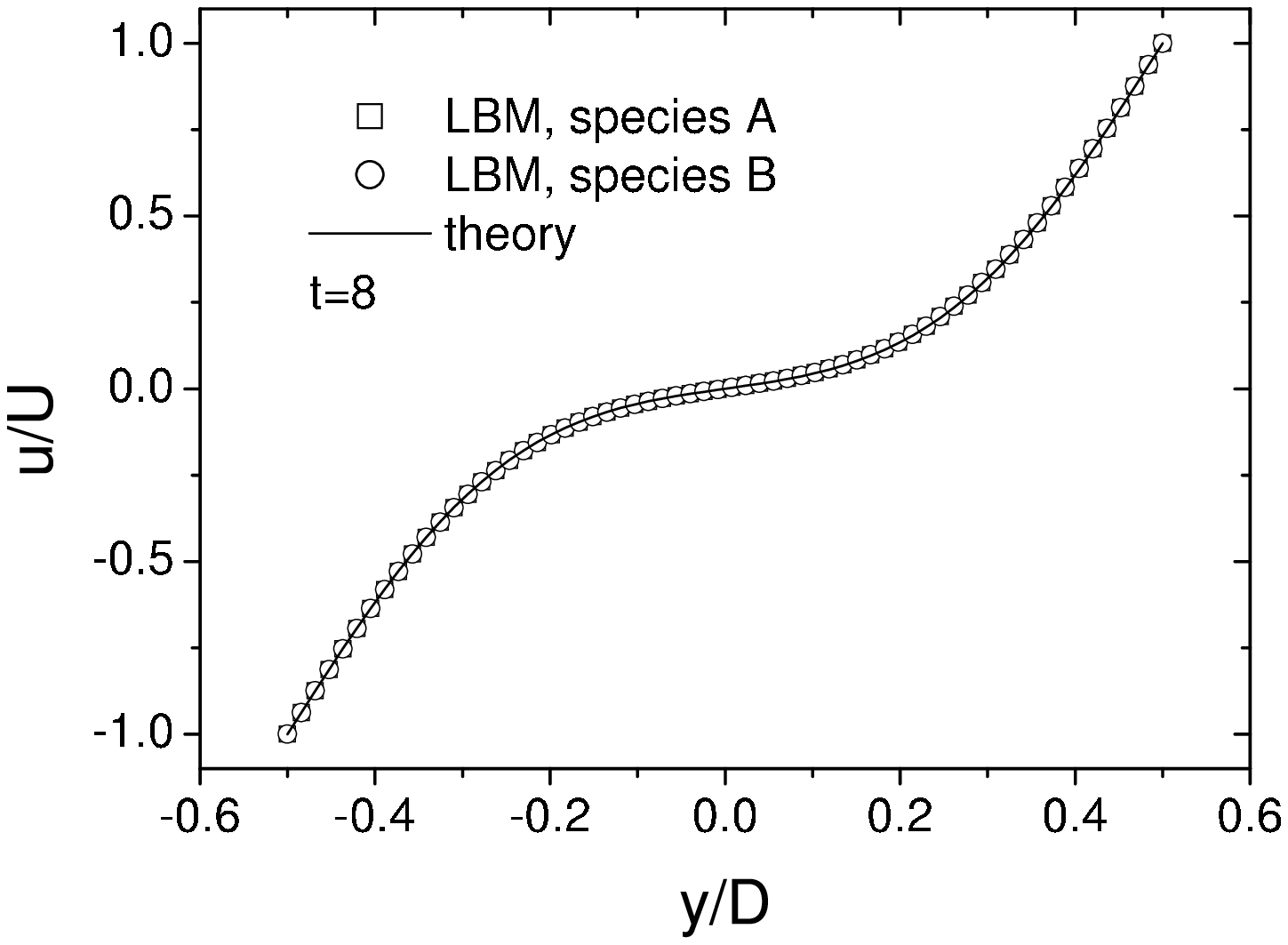}{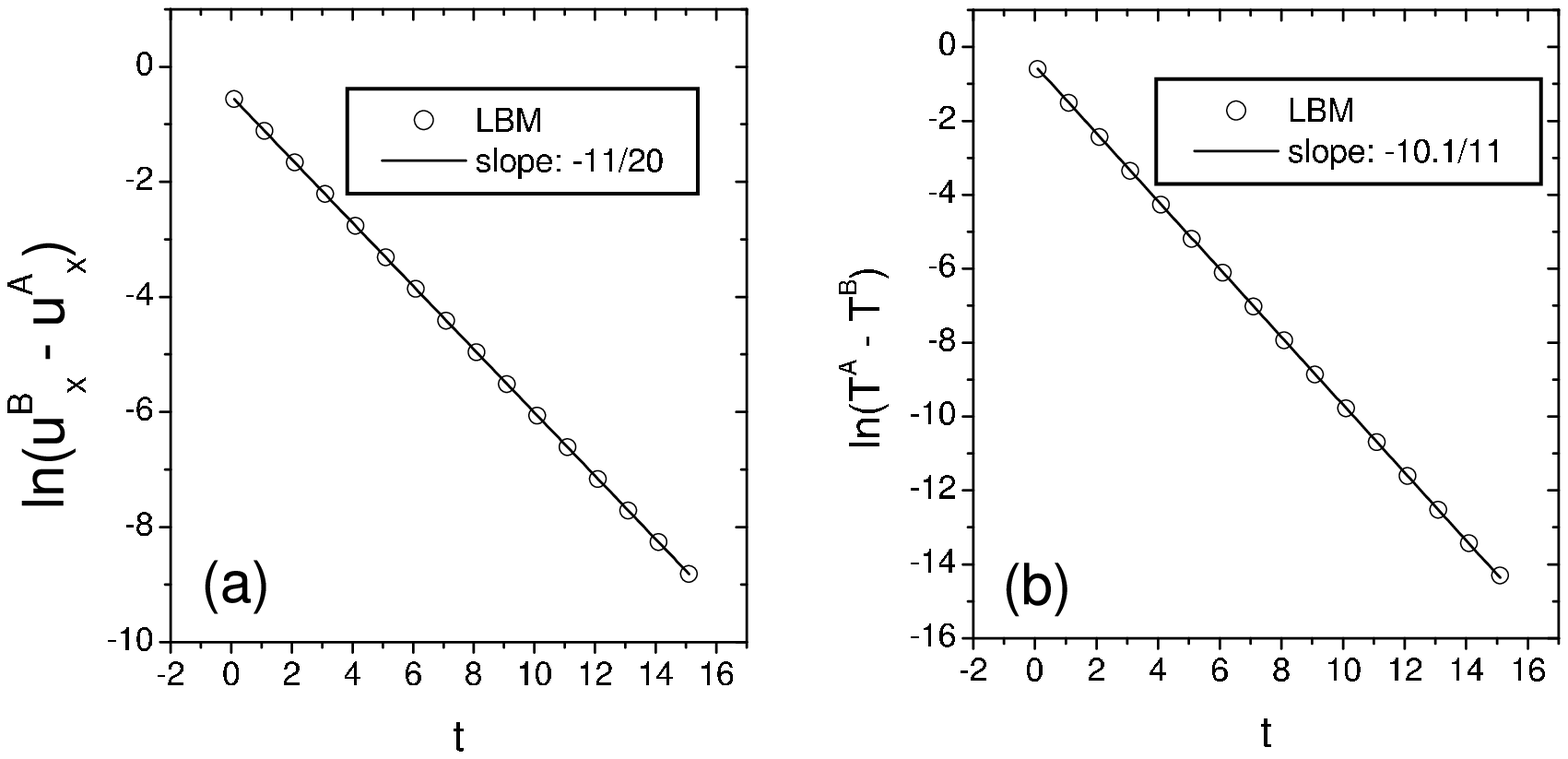}
\caption{ Horizontal velocity profiles along a vertical line for the 
two species, $A$ and $B$, at time $t=8$. The symbols are for simulation 
results. The solid line corresponds to the theoretical result,
Eq. (\ref{uy}).
Parameters used in the two-fluid FDLBM are $m^A=m^B=1$, $T=1$,
 $n^A=n^B=1$, $\gamma =0.001$%
, $\tau^{AA}=\tau^{BB} =\tau^{AB}=\tau^{BA}=0.2$. Parameters used in 
Eq. (\ref{uy}) are $\eta =\eta^A=0.1$, $\rho =\rho^A=1$. 
}
\label{Fig1}
\caption{Uniform relaxation processes. (a) Equilibration of velocities;
(b) Equilibration of temperatures. The symbols are for simulation
results. The solid lines possess the theoretical slopes. 
Common parameters for the simulations in (a) and (b) are 
$n^A=10$, $n^B=1$, $m^A=1$, $m^B=10$, 
$\tau^{AA}=\tau^{BB}=1$, $\tau^{AB}=10$, $\tau^{BA}=1$. In (a) the
 initial conditions are 
$u^{A(0)}_x=-u^{B(0)}_x=-0.3 $, $u^{A(0)}_y=u^{B(0)}_y=0 $, and
 $T^{A(0)}=1.3$, $T^{B(0)}=0.7$.
 The slope of the solid line in (a) is $-11/20$, which is consistent
 with Eq. (\ref{vrelax}). 
In (b) the initial conditions are
$\mathbf{u}^{A(0)}=\mathbf{u}^{B(0)}=0 $, and
 $T^{A(0)}=1.3$, $T^{B(0)}=0.7$. 
The slope of the solid line in (b) is $-10.1/11$, 
which is consistent with the first term of right-hand side of 
Eq.(\ref{Trelax}).
The second superscript ``(0)'' denotes the corresponding initial value. 
This figure shows an example where the particle masses of the two
 species are significantly different.}
\label{Fig2}
\end{figure}


The second one is the uniform relaxation process, which is an ideal process
to indicate the equilibration behavior of the mixture\cite{XuFDLBM1}. By
neglecting the force terms and terms in spatial derivatives, the
Navier-Stokes equations (\ref{mass})-(\ref{energy}) give
\begin{equation}
\frac \partial {\partial t}\rho ^A=0\mathtt{,}  \label{rho}
\end{equation}
\begin{equation}
\frac \partial {\partial t}\left( \mathbf{u}^B-\mathbf{u}^A\right) =-\frac
1\rho \left( \frac{\rho ^A}{\tau ^{BA}}+\frac{\rho ^B}{\tau ^{AB}}\right)
\left( \mathbf{u}^B-\mathbf{u}^A\right) \mathtt{,}  \label{vrelax}
\end{equation}
\begin{equation}
\frac{\partial \left( T^B-T^A\right) }{\partial t}=-\frac 1n\left( \frac{n^A%
}{\tau ^{BA}}+\frac{n^B}{\tau ^{AB}}\right) \left( T^B-T^A\right) +\frac{%
\rho ^A\rho ^B}{2k_Bn\rho }\left( \frac 1{\tau ^{AB}}-\frac 1{\tau
^{BA}}\right) \left( \mathbf{u}^B-\mathbf{u}^A\right) ^2\mathtt{.}
\label{Trelax}
\end{equation}
The flow velocities of the two components equilibrate exponentially with
time. (For example, see Fig. 2(a).) The equilibration of flow velocities also affects that of the
temperatures. When the flow velocity difference is zero, the temperatures
equilibrate exponentially with time. (For example, see Fig. 2(b).)
The simulation results agree well with Eqs. (\ref{vrelax}) and (\ref
{Trelax}). 

\section{Conclusions and remarks}

The Chapman-Enskog analysis shows what properties the discrete
Maxwellian distribution function $f_{ki}^{A\left( 0\right) }$ should follow.
Those requirements tell the lowest order of the flow velocity
$\mathbf{u}^A$ in the Taylor expansion of $f_{ki}^{A\left( 0\right) }$.
The highest rank of tensors of the particle velocity 
$\mathbf{v}^A$  
in the requirements on the truncated 
$f_{ki}^{A\left( 0\right) }$ determines the needed isotropy
of the DVM. The incorporation of the force terms makes no additional
requirement on the isotropy of the DVM. The present approach works for 
binary neutral fluid mixtures. One possibility to introduce
interfacial tension is to modify the pressure tensors\cite{EPL32463}, 
which is implemented by changing the force terms\cite{SofoneaPRE}. 
The specific force terms or pressure tensors, which are out of the scope of 
this Letter,
depend on the system under consideration, but can be resolved under the same two-dimensional
61-velocity model (D2V61). For
binary fluids with disparate-mass components, say $m^A\ll m^B$, only if the
total masses and temperatures of the two species are not significantly
different, does Sirovich's kinetic theory works\cite{XuFDLBM1}, so do the
corresponding FDLBMs. (See Fig. 2 for an example.) When the masses and/or the temperatures of the two
components are greatly different, the two-fluid kinetic theory 
should be modified. In those cases, the Navier-Stokes equations and the
FDLBMs are not
symmetric about the two components, but the FDLBMs can still be resolved
under the D2V61 model. The formulation procedure is straightforward. 
In practical simulations, numerical errors from the finite-difference
schemes should be quantified.

\acknowledgments
The author thanks Prof. G. Gonnella for guiding him into the LBM field
and Profs. H. Hayakawa, V. Sofonea, S. Succi, M. Watari for helpful
discussions. This work
is partially supported by Grant-in-Aids for Scientific Research (Grant No.
15540393) and for the 21-th Century COE ``Center for Diversity and
Universality in Physics'' from the Minstry of Education, Culture and Sports,
Science and Technology (MEXT) of Japan.

\end{document}